\documentclass[aps,twocolumn]{revtex4}
\usepackage{graphicx}
\usepackage{amsmath,amssymb}

\begin{document}
\title{Myosin V passing over Arp2/3 junctions: branching ratio calculated from
  the elastic lever arm model} 

\author{Andrej Vilfan}
\affiliation{J. Stefan Institute, Jamova 39, 1000 Ljubljana,
    Slovenia} 
\email{andrej.vilfan@ijs.si}
\date{\today}

\begin{abstract}
Myosin V is a two-headed processive motor protein that walks
  in a hand-over-hand fashion along actin filaments. When it encounters a
  filament branch, formed by the Arp2/3 complex, it can either stay on the
  straight mother filament, or switch to the daughter filament. We study both
  probabilities using the elastic lever arm model for myosin V.  We calculate
  the shapes and bending energies of all relevant configurations in which the
  trail head is bound to the actin filament before Arp2/3 and the lead head is
  bound either to the mother or to the daughter filament.  Based on the
  assumption that the probability for a head to bind to a certain actin subunit
  is proportional to the Boltzmann factor obtained from the elastic energy, we
  calculate the mother/daughter filament branching ratio.  Our model predicts a
  value of 27\% for the daughter and 73\% for the mother filament.  This result
  is in good agreement with recent experimental data.
\end{abstract}

\maketitle

\section{Introduction}

Myosin V is a two-headed processive motor protein from the myosin superfamily,
involved in different forms of intracellular transport
\cite{Vale2003b,Sellers.Veigel2006}.  It has drawn a lot of attention in recent
years and is now one of the best studied motor proteins.  The experiments have
characterized it mechanically
\cite{Mehta.Cheney1999,Rock.Spudich2000,Rief.Spudich2000,Veigel.Molloy2002,Purcell.Sweeney2002,Clemen.Rief2005,Gebhardt.Rief2006,Uemura.Ishiwata2004},
biochemically
\cite{De_La_Cruz.Sweeney1999,De_La_Cruz.Ostap2000a,De_La_Cruz.Ostap2000b,Yengo.Sweeney2002},
optically \cite{Ali.Ishiwata2002,Forkey.Goldman2003,Yildiz.Selvin2003} and
structurally
\cite{Walker.Knight2000,Burgess.Trinick2002,Wang.Sellers2003,Coureux.Houdusse2003}.
These studies have shown that myosin V walks along actin filaments in a
hand-over-hand fashion \cite{Yildiz.Selvin2003,Warshaw.Trybus2005} with an
average step size of about 35 nm, roughly corresponding to the periodicity of
actin filaments
\cite{Mehta.Cheney1999,Rief.Spudich2000,Walker.Knight2000,Veigel.Molloy2002,Ali.Ishiwata2002},
a stall force of around 2 pN \cite{Rief.Spudich2000} and a run length of a few
microns \cite{Rief.Spudich2000,Sakamoto.Sellers2003,Baker.Warshaw2004}.  Under
physiological conditions, ADP release has been identified as the time limiting
step in the duty cycle \cite{De_La_Cruz.Sweeney1999,Rief.Spudich2000}.

The Arp2/3 complex \cite{Pantaloni.Carlier2000} initiates actin filament
branching in the vicinity of a protruding edge of a cell. The complex consists
of 7 subunits (Arp2, Arp3 and ARPC1 through ARPC5) and is activated by
WASp/Scar proteins \cite{Machesky.Pollard1999}.  It binds to the side of one
(``mother'') actin filament and initiates the nucleation of a second
(``daughter'' filament), which starts growing with the fast growing end (``+''
end) away from the Arp2/3 complex. The mother and the daughter filament enclose
an angle of $70^\circ$.

The question what happens to a myosin V motor when it arrives at an Arp2/3
mediated actin filament junction is of interest for several reasons.  First,
the branching behavior is important for understanding vesicle transport in the
actin cortex.  And second, it is of high interest when studying the fundamental
mechanism of myosin V stepping, because it represents a well defined situation
in which predictions from different theoretical models can be tested against
the experimental result.
Our aim in this article is to use the elastic lever arm model for myosin V,
which is described in detail in a previous article \cite{Vilfan2005}, to
predict the dynamics of a myosin V motor that passes over an Arp2/3 junction.
In particular, we will calculate the probabilities that a motor continues along
the mother or the daughter filament.

\section{Model}

The idea behind the elastic lever arm model for myosin V is to describe the
dimeric motor as an assembly of two identical heads, connected together and to
the cargo-binding tail with elastic lever arms. The model allows us to derive
the properties of a dimeric molecule, such as step size distribution,
force-velocity relation and processivity, from the properties of an individual
head, such as geometry, chemical kinetics and elasticity.  In this respect the
approach is different from the class of discrete stochastic models, which
describe the motor as a single unit
\cite{Kolomeisky.Fisher2003,Skau.Turner2006}.

We describe each head with a 5-state mechano-chemical model, similar to that
for muscle myosin (e.g., \cite{vilfan2003b}), where each state (with bound
ADP.Pi, ADP (pre-powerstroke), ADP (post-powerstroke), without a nucleotide,
detached) is connected with a certain orientation of the lever arm, as
determined with electron microscopy (EM)
\cite{Walker.Knight2000,Burgess.Trinick2002}.  The lever arms are modeled as
elastic beams, connected with a flexible joint.  A recent study measuring
fluctuations in the position of the free head \cite{Dunn.Spudich2007} has
demonstrated that the lead head diffuses around the joint freely before binding
to the next actin site, meaning that there is no detectable elastic energy cost
connected with variation in the angle between the two lever arms.  The very
nature of protein flexibility, which mainly originates from the twisting of
bonds between carbon atoms in the backbone, leads us to the conclusion that the
joint is also fully flexible with regard to rotation of each lever arm along
its axis.  Similar flexibility has also been observed in myosin II
\cite{Knight.Trinick1984,Offer.Knight1996,Knight1996}.

The calculation of the branching probability is simplified a lot if we make the
following assumptions.  First, we assume that the binding of the lead head
always leads to a step, which means that its unbinding is significantly slower
than the step that follows in the regular cycle (Pi release). Second, we assume
that the probability for the lead head to bind to site $j$ if the trail head is
bound to site $i$ is given by the Boltzmann factor
\begin{equation}
  \label{eq:boltzmann}
  P_{j|i}=\frac{\exp\left( - \frac {U_{i,j}}{k_BT} \right) }{
\sum_{j'} \exp\left( - \frac {U_{i,j'}}{k_BT} \right) }\;.
\end{equation}
$U_{i,j}$ denotes the elastic energy of deformed lever arms when the trail head
is in the post-powerstroke state, bound to site $i$, and the lead head in the
pre-powerstroke state to site $j$.  In the following, we will use the notation
where the sites on the mother filament are marked with $({\rm M},i)$ and those
on the daughter filament with $({\rm D},i)$.  For example, $P_{{\rm D},2|{\rm
    M},-9}$ denotes the conditional probability for the lead head to bind to
site $2$ on the daughter filament if the trail head is bound to the site $-9$
on the mother filament. We enumerate the actin subunits so that the central
subunit under the Arp2/3 complex on the mother filament has the index 0.
Positive indices denote subunits towards the "+" end and negative towards the
"-" end.  Subunits of the daughter filament are enumerated from 0 onwards.
Note that sites numbered -2, 0 and 2 on the mother filament are not accessible
for a myosin V head, because of steric hindrance with the Arp2/3 complex.

\begin{figure}
  \begin{center}
   \includegraphics{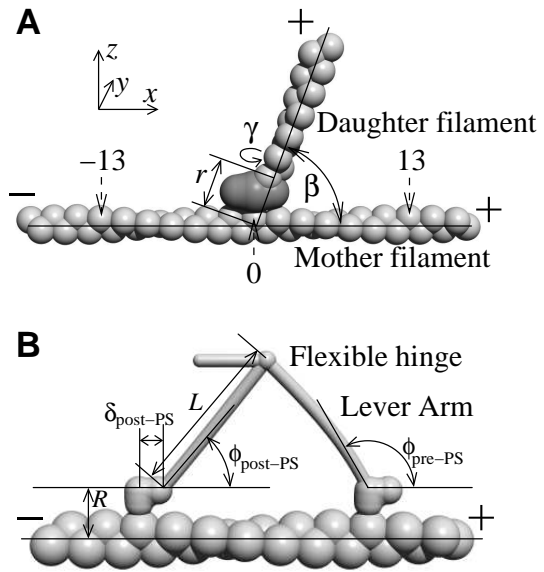}\\
  \end{center}
\caption{Arp2/3 junction and a dimeric myosin V motor.  (A) The Arp2/3 complex
  is attached to the side of the mother filament occupying subunits -2, 0 and
  2.  It nucleates the growth of a daughter filament, whose position is
  determined by the angles $\beta$ (branching angle), $\gamma$ (twist) and $r$
  (distance from the first subunit to the center of the mother filament).
  (B) The myosin V motor is consists of two heads, connected with lever arms,
  which we describe as elastic rods. The proximal end of each lever arm always
  leaves the head at a fixed angle $\phi$, which depends on the nucleotide
  state of that head.  The distal ends of both lever arms are connected with a
  fully flexible joint.}  
\label{fig:geom}
\end{figure}
\begin{table}
    \caption{Geometric parameters of the Arp2/3 junction and a myosin V head.}
    \label{tab:geometric}
    \begin{center}
      \begin{tabular}{lll}
        \hline
        Distance actin subunits & $a$ & $2.75\,{\rm nm}$\\
        Angle actin subunits & $\theta_0$ & $167.14^\circ$\\
        Daughter filament angle & $\beta$ & $70^\circ$\\
        Daughter filament offset & $r$ & $12\,{\rm nm}$\\
        Daughter filament rotation & $\gamma$ & $39^\circ$\\
        Lever arm start:  radial pos. & $R$ & $8\,{\rm nm}$\\
        Lever arm start: displacement & $\delta_{\rm pre-PS}$ & 0 \\
        & $\delta_{\rm post-PS}$ & $3.5\,{\rm nm}$ \\
        Lever arm angle &  $\phi_{\rm pre-PS}$ & $115^\circ$ \\
        & $\phi_{\rm post-PS}$ &   $50^\circ$ \\
        Lever arm length & $L$ & $26\,{\rm nm}$\\
        \hline
      \end{tabular}
    \end{center}
\end{table}

The structure of the Arp2/3 complex and both actin filaments
(Fig.~\ref{fig:geom}) has been determined from EM studies
\cite{Volkmann.Hanein2001,Egile.Hanein2005} and its parameters are summarized
in Table \ref{tab:geometric}.  While we approximated actin with the commonly
assumed 13/6 helix in the original paper \cite{Vilfan2005}, we use a more
accurate 28/13 helix here, with the angle $\theta_0=167.14^\circ$ between
adjacent subunits.  A detailed discussion on different helix models and their
consequence for the calculated step size distribution can be found in
\cite{vilfan2005b}.

When a head is bound to the site $i$, the starting point and the unit vector
giving the initial direction of the lever arm are given by
\begin{equation}
  \label{eq:starting1}
  {\bf x}_0={\bf R}_x(-i\theta_0) 
\left(\!\!\! \begin{array}{c}
ia +\delta \\
0\\
R
\end{array}\!\!\!
\right)
\quad
 \hat {\bf t}_0={\bf R}_x(-i\theta_0)
\left(\!\!\!\begin{array}{c}
\cos \phi_i \\
0\\
\sin \phi_i
\end{array}\!\!\!
\right)
\end{equation}
where ${\bf R}_x$ denotes the rotation matrix around the $x$ axis,
\begin{equation}
  \label{eq:rx}
{\bf R}_x(\theta)= \left( \begin{array}{ccc} 1 & 0 & 0 \\
0 & \cos \theta  & -\sin \theta \\
0 & \sin \theta  & \cos  \theta
\end{array}
\right) 
\end{equation}

  For a head
bound to the daughter filament, the two vectors read
\begin{align}
  {\bf x}_0 &={\bf R}_y(-\beta) {\bf R}_x(\gamma-i\theta_0)
\left(\!\!\! \begin{array}{c}
r+ia +\delta \\
0\\
R
\end{array}\!\!\!
\right)\nonumber
\\
  \label{eq:starting2}
 \hat {\bf t}_0 &={\bf R}_y(-\beta) {\bf R}_x(\gamma-i\theta_0)
\left( \!\!\!\begin{array}{c}
\cos \phi_i \\
0\\
\sin \phi_i
\end{array}\!\!\!
\right)
\end{align}
with
\begin{equation}
  \label{eq:ry}
{\bf R}_y(\beta)=
\left(\begin{array}{ccc} 
\cos \beta & 0  & \sin \beta \\
0 & 1 & 0 \\
 -\sin \beta & 0 & \cos \beta 
\end{array}\right)\;.
\end{equation}
Here $\beta=70^\circ$ denotes the angle between the mother and the daughter
filament and $\gamma=39^\circ$ the rotation of the daughter filament around its
axis (see Fig.~\ref{fig:geom}).

We calculate the shapes of both lever arms as described in
Ref.~\cite{Vilfan2005} by minimizing the bending energy $U=\int ds \, EI
(C(s))^2/2$, where $C(s)$ denotes local curvature.  For the bending modulus of
the lever arm we use the value $EI=1500\,{\rm pN\,nm^2}$, which corresponds to
a ``spring constant'' of $\kappa=3EI/L^3=0.25\,{\rm pN/nm}$, measured at the
tip of the lever arm.  This value was originally estimated from the stall
force, but it shows good agreement with direct optical tweezer measurements
\cite{Veigel.Sellers2005}.  We neglect any additional compliance resulting from
the head or converter domain. Because most of the bending takes place in the
proximal part of the lever arm, we expect that its effect would not be
significantly different. For the temperature we use the value $T=27^\circ \rm
C$.

\section{Results}

\begin{figure*}[p]
\begin{center}  
\includegraphics{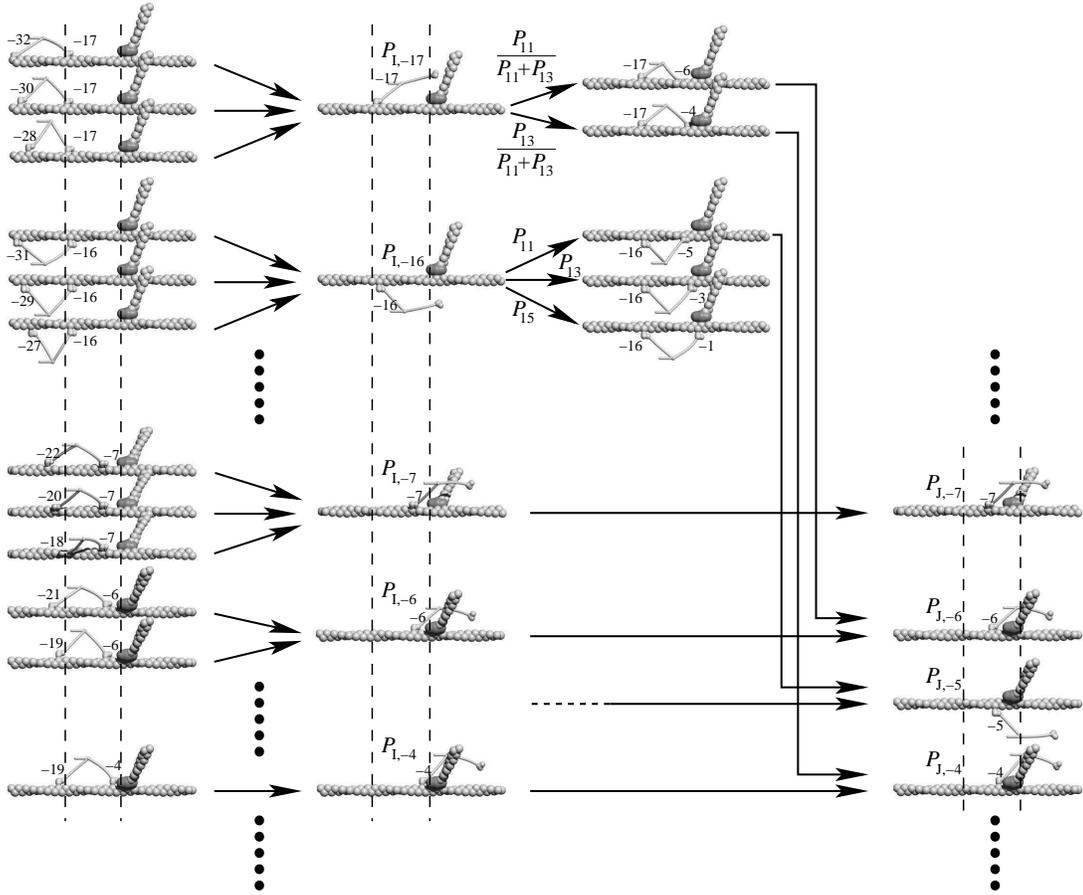}
\end{center}
\caption{Different pathways on which a myosin V motor can approach the Arp2/3
  junction. The probability distributions for the first accessed site in the
  two intervals marked with dashed lines are denoted as $P_{{\rm I},i}$ and
  $P_{{\rm J},i}$.  $P_{{\rm I},i}$ represents the probability that $i$ is the
  first accessed site with $-17\le i \le -3$ (between dashed lines in the first
  or second column).  For values between $-17$ and $-7$, this state can be
  reached with 3 different step sizes. For $i=-6$ and $i=-5$, it can be reached
  with 2 different step sizes (13 and 15). If it is reached by a shorter step
  (11 subunits), it means that the preceding binding site was already inside
  the interval $-17\ldots -3$, so $i$ is not the first accessed site within it.
  For $i=-4$ and $i=-3$, the site can only be reached with a 15-subunit step.
  $P_{{\rm J},i}$ denotes the probability that $i$ is the first accessed
  subunit in the interval $-13\le i \le 1$ (dashed lines in the right column).}  
  \label{fig:cartoon}
\end{figure*}

\begin{figure*}[p]
  \includegraphics{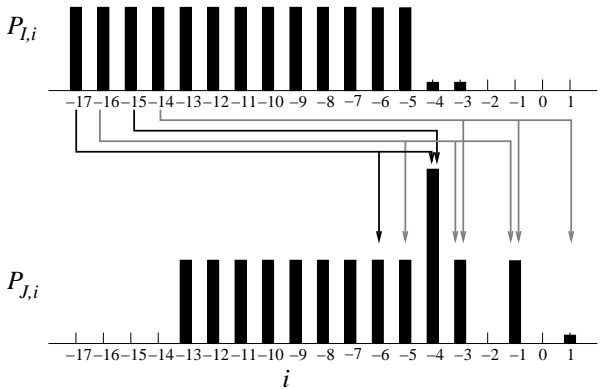}\hfill
\begin{minipage}[b]{\columnwidth}
\caption{Probability distribution $P_{{\rm I},i}$ (upper diagram) for the first
  accessed binding site on the mother filament in the interval $-17\le i \le
  -3$.  The lower diagram shows the probability distribution $P_{{\rm J},i}$
  for the first accessed site with $-13\le i \le 1$. $P_{{\rm J},i}$ is
  calculated according to Eq.~(\ref{eq:pt}), by redistributing the
  probabilities for $-17$ to $-14$ to other sites, as indicated by arrows.  For
  example, if -17 is the first accessed site with $-17\le i \le -3$, the first
  accessed site with $-13 \le i \le 1$ can either be -6 or -4. Note that the
  total probability that the motor binds to site -4 is higher than for any
  other site, which is due to the inaccessibility of site -2.}  
\label{fig:probs}
\end{minipage}
\end{figure*}

\subsection*{Stepping of myosin V in the absence of the Arp2/3 complex}
When a myosin V motor is sufficiently far away from the Arp2/3 complex, it
exhibits the stepping pattern that has already been discussed in
Refs.~\cite{Vilfan2005,vilfan2005b}.   We restrict the step lengths of an
unperturbed motor to 11, 13 and 15 subunits, with probabilities
\begin{align}
P_{11}&=P_{{\rm M},i+11|{\rm M},i}\approx 0.005\;,\\
P_{13}&=P_{{\rm M},i+13|{\rm M},i}\approx0.895\;,\\
P_{15}&=P_{{\rm M},i+15|{\rm M},i}\approx0.1\;,
\end{align}
calculated from Eq.~(\ref{eq:boltzmann}).  Probabilities for other step sizes,
such as $P_9$ and 
 $P_{17}$  turn out to be very small.  Note that  in this
calculation  $P_{11}$ and $P_{15}$ are likely to be   somewhat  underestimated -
their values are somewhat higher if we take into account torsional fluctuations
in the actin helix \cite{vilfan2005b}. 

 The average step size can be calculated from these probabilities as
\begin{equation}
\bar l=11 P_{11}+13 P_{13}+15 P_{15}\approx 13.2\;.
\end{equation}
In the absence of the Arp2/3 junction, the fraction of sites that get accessed
by a passing myosin V motor is $1/\bar l$.  This is also the probability that a
site before the junction ($i\le -7$, as will be shown later) ever gets accessed
by the motor: 
\begin{equation}
  \label{eq:prob0}
  P_{{\rm M},i}=\frac 1 {\bar l}\qquad  \text{for } i\le -7\;.
\end{equation}

\begin{table}
  \caption{ Probability distribution ${P_{{\rm I},i}}$ for the first accessed binding
    site in the interval $-17\le i \le -3$.   ${P_{{\rm J},i}}$ denotes the
    probability distribution for the first accessed site in the interval $-13
      \le i \le 1$. ${P_{{\rm J},i}}$ is calculated according to Eq.~(\ref{eq:pt}).}
\label{tab:probs}
\begin{center}

\begin{tabular}{llll}
\hline
$i$ & $P_{{\rm I},i}$ & $P_{{\rm J},i}$&$P_{{\rm J},i}$  \\
\hline
-17 &$ 1 /{\bar l}$ & & \\
 -16 &$1 /{\bar l}$ &   & \\
 -15 &$1 /{\bar l}$ &   & \\
 -14 &$1 /{\bar l}$ &   & \\
 -13&$1 /{\bar l}$ & $1 /{\bar l}$ & 0.0758\\
 -12 & $1 /{\bar l}$ &$1 /{\bar l}$ & 0.0758\\
 $\vdots$ & $\vdots$ &$\vdots$ & \vdots \\
 -7 & $1/{\bar l}$ & $1/{\bar l}$ & 0.0758\\
 -6 & $(P_{13}+P_{15})/{\bar l}$ &
 $(P_{13}+P_{15}+\frac{P_{11}}{P_{11}+P_{13}})/{\bar l}$ & 0.0759
 \\
 -5 &   $(P_{13}+P_{15})/{\bar l} $ &   $1/{\bar l} $ & 0.0758 \\
 -4 &   $P_{15} /{\bar l}$&   $
 (P_{15}+\frac{P_{13}}{P_{11}+P_{13}}+1)/{\bar l}  $ & 0.1587\\
 -3 & $P_{15}/{\bar l} $ & $1/ {\bar l} $ & 0.0758\\ 
 -2 &  & - \\
 -1 &  &  $({P_{15}}+{P_{13}})/{\bar l}$ & 0.0754\\
 0 &  & - \\
 1 &  &  ${P_{15}}/{\bar l}$ & 0.0075\\
\hline
\end{tabular}
\end{center}
\end{table}

\subsection*{Initial state: First accessed site in the interval $\mathbf{ -17\le i \le -3}$}

We start our analysis at the point where a head of an approaching myosin V
first passes the subunit $-17$ or binds to it. This way the initial state is
definitely not influenced by the presence of the Arp2/3 complex.  By counting
only the first accessed site in this interval, we avoid double-counting of
events where the motor binds, for example, first to site -17 and then to -4.

We denote with $P_{{\rm I},i}$ the probability that once the first head has
bound to any subunit in the interval $-17\le i \le -3$, this has happened at
subunit $i$.  $P_{{\rm I},i}$ can be calculated in a way that is illustrated in
Fig.~\ref{fig:cartoon}.  The sites between -17 and -7 can only be reached from
outside the interval.  Therefore, whenever the motor binds to one of them, it
becomes the first accessed site in this interval.  The probability $P_{{\rm
    I},i}$ then equals the probability that the site $i$ ever gets accessed by
the motor, which is $1/\bar l$, according to Eq. (\ref{eq:prob0}).  For $i=-6$
and higher the situation becomes different. Site -6 counts as the first site in
this interval if the preceding step size is 13 or 15 subunits, but not if it is
11.  Therefore, the corresponding probability is $P_{{\rm
    I},-6}=(P_{13}+P_{15})/\bar l$.  The probability that the first site
accessed in this interval is -5, $P_{{\rm I},-5}$, has the same value.
Finally, the sites -4 and -3 will be the first accessed sites in the interval
if they are following a 15 subunit step.  Therefore, their probabilities are
$P_{{\rm I},-4}=P_{{\rm I},-3}=P_{15}/\bar l$.  The probabilities $P_{{\rm
    I},i}$ are given in the second column of Table \ref{tab:probs} and shown in
the top graph of Fig.~\ref{fig:probs}.

\begin{figure*}
  \begin{center}
    \includegraphics{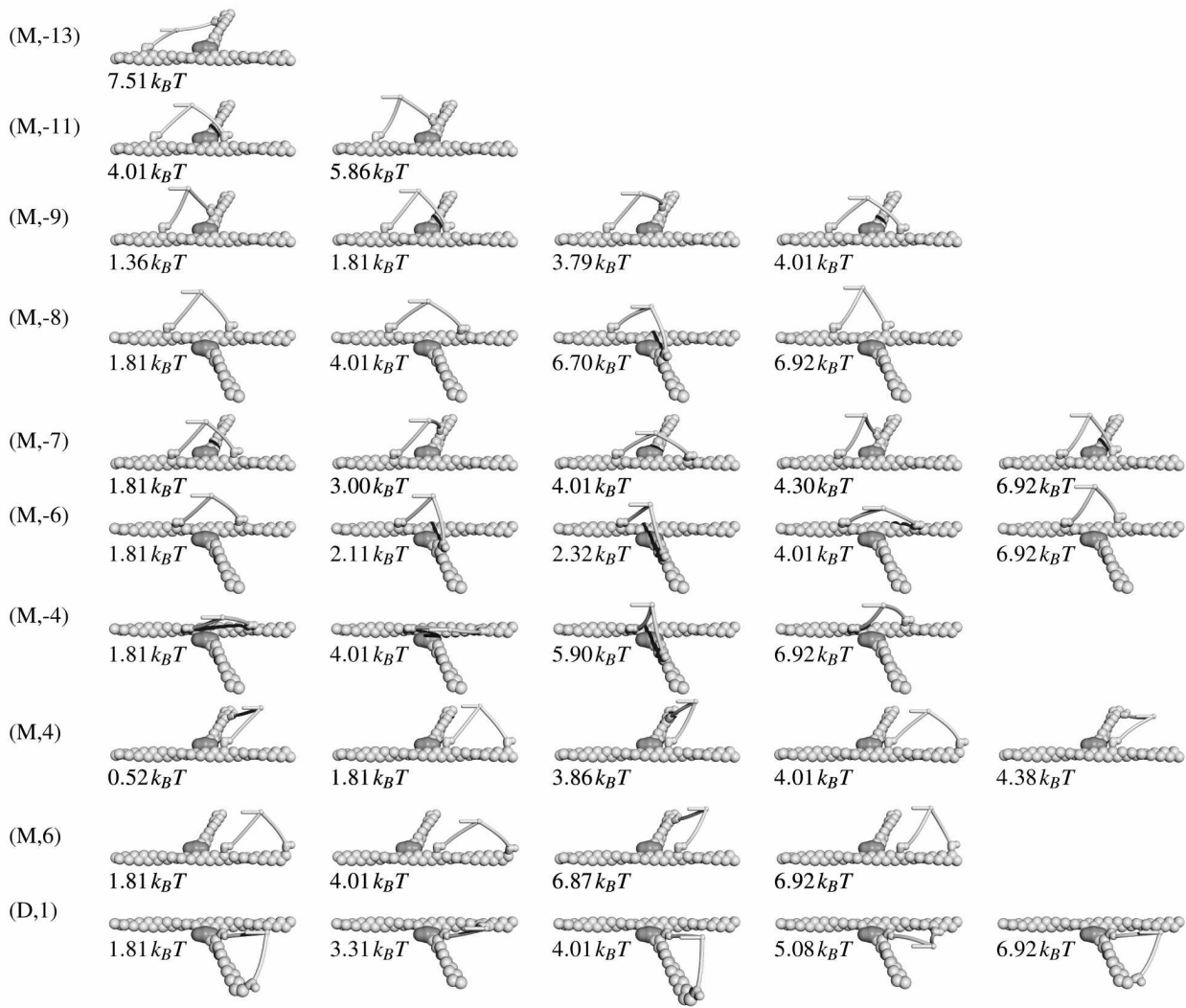}
  \end{center}
\caption{Most probable lead head binding sites for different trail head
  positions. For selected trail head binding sites ${i}$, a collection of lead
  head binding sites and their bending energies ${U_{i,j}}$ (in units of
  ${k_BT}$) is shown.  Some trail head positions which do not have significant
  branching probabilities (e.g., $-12$) are omitted.  The shape of both lever
  arms in each configuration is calculated numerically, by minimizing the
  bending energy.}  
\label{fig:gallery}
\end{figure*}

As a next step, we will derive the probability distribution $P_{{\rm J},i}$ for
the first accessed site in the interval $-13\le i \le 1$.  We choose this
interval because (M,-13) is the first site from which the motor can reach the
daughter filament.  This distribution can be obtained from $P_{{\rm I},i}$ by
redistributing probabilities for sites between -17 and -14 according to the
conditional probability for the next step
\begin{equation}
  \label{eq:pt}
  P_{{\rm J},j}=P_{{\rm I},j}+\sum_{i=-17}^{-14}P_{{\rm M},j|{\rm M},i} P_{{\rm I},i}\;,
\end{equation}
as shown in Fig.~\ref{fig:probs}.  The values of $P_{{\rm J},j}$ are given in the
third and fourth column of Table \ref{tab:probs}. This distribution defines the
state from which we will calculate the branching ratio at the Arp2/3 junction
in the following section.

\subsection*{Conditional branching ratio}

Now we can calculate the conditional probabilities that the lead head binds to
the daughter filament, if the trail head is bound to a mother filament subunit
$i$ fulfilling $-13\le i \le 1$.  We denote this probability as $P_{d|{\rm
    M},i}$.

\begin{figure*}
  \begin{center}
  \includegraphics{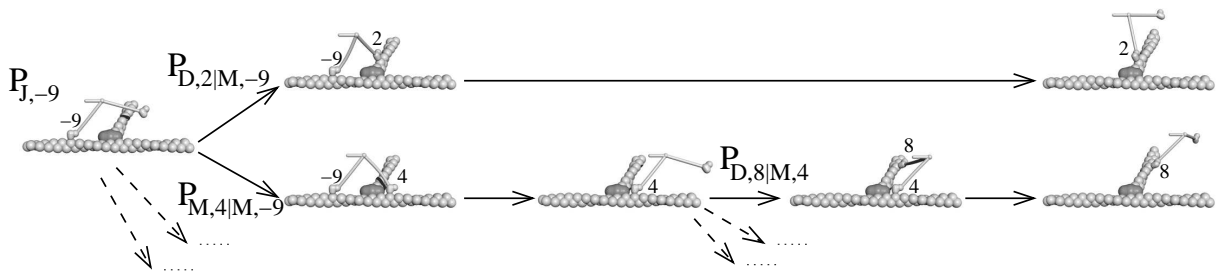}    
  \end{center}
 \caption{An example of a one-step (upper path) and two-step (lower path)
  process through which the myosin V motor can reach the daughter filament. The
  probability that the trail head is bound to subunit (M,-9) in the initial
  state is $P_{{\rm J},-9}$.  In the next step, the lead head can bind to site
  (D,2) with conditional probability $P_{{\rm D},2|{\rm M},-9}$. We denote such
  processes as \textit{one-step}. Alternatively, the lead head can bind to the
  site (M,4) with conditional probability $P_{{\rm M},4|{\rm M},-9}$ and,
  subsequently, the other head can bind to site (D,8) with conditional
  probability $P_{{\rm D},8|{\rm M},4}$. This is an example of a
  \textit{two-step} process.  Note that this scheme shows just two examples of
  possible pathways and omits alternatives that are indicated by dashed
  arrows.}
 \label{fig:onevstwostep}
\end{figure*}

\begin{figure}
  \begin{center}
    \includegraphics{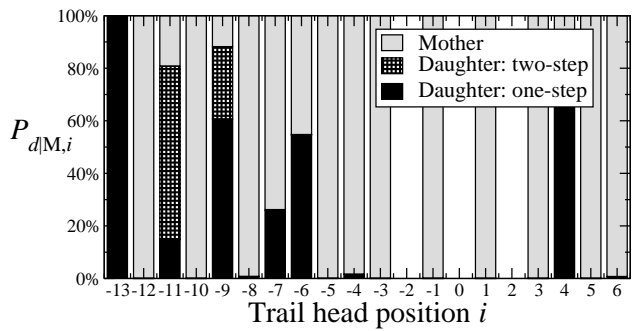}
  \end{center}
 \caption{Probability that the motor will step on and continue its walk along
  the daughter filament if the trail head is initially bound to site $i$ on the
  mother filament, $P_{d|{\rm M},i}$ (black and hatched bars), as calculated
  from Eq.~(\ref{eq:sidebr0}). The black bars show the contribution of one-step
  processes, in which the lead head binds to a site on the daughter filament
  immediately.  The hatched bars show the contribution of two-step processes in
  which the motor first binds to another site on the mother filament (usually
  4) and then in the second step to the daughter filament.  The grey bars show
  the probability that the motor continues along the mother filament.}  
\label{fig:table}
\end{figure}

Figure~\ref{fig:gallery} shows the most relevant dimer configurations with the
trail head on mother filament sites between -15 and 6 and the lead head either
on mother or on daughter filament.  For each trail head position, the
conditional probability that the lead head binds to a certain site is given by
Eq.~(\ref{eq:boltzmann}) with the index $j'$ running over all accessible
mother- as well as daughter-filament sites -- corresponding to one row in
Fig.~\ref{fig:gallery}.  Generally speaking, the daughter filament can either
be reached directly, in a one-step process such as (M,-6)$\to$(D,3), or in a
two-step process such as (M,-9)$\to$(M,4)$\to$(D,8).  Two examples are shown in
Fig.~\ref{fig:onevstwostep}.  The probability that a motor with the trail head
bound to site (M,$i$) eventually binds to the daughter filament is
\begin{equation}
  \label{eq:sidebr0}
  P_{d|{\rm M},i}=\sum_{j} \left( P_{{\rm D},j|{\rm M},i}
    +\sum_{k}  P_{{\rm D},j|{\rm M},k}  P_{{\rm M},k|{\rm M},i}  \right)\;.
\end{equation}
Here the first term denotes one-step processes and the second term two-step
processes, where the motor first binds to site $k$ on the mother filament and
subsequently to site $j$ on daughter filament.  Our numerical calculation shows
that the only significant terms are those with $k=4$.  The branching ratio
$P_{d|{\rm M},i}$ for each trail head position is shown in
Fig.~\ref{fig:table}.  The graph shows separately the contributions of one- and
two-step processes.

\subsection*{Total branching ratio}

With these probabilities and weights $P_{{\rm J},i}$ we finally obtain the
branching ratio for the daughter filament:
\begin{equation}
  \label{eq:sidebr}
  P_d=\sum_{i=-13}^{1} P_{{\rm J},i} P_{d|{\rm M},i} \approx 0.27\;.
\end{equation}
If one head binds to the daughter filament, there is still some probability
that the next head will bind back to the mother filament.  One such example,
with the trail head on the site (D,1), is shown in the last row in
Fig.~\ref{fig:gallery}.  However, the contribution of such events to the total
branching ratio is not significant.

\section{Discussion}

While the exact result calculated above does require to take into account all
the individual configurations, its order of magnitude can also be understood
with a simple ``handwaving'' argument, which goes as follows.  Roughly
speaking, the approaching myosin V motor can either reach the Arp2/3 complex on
the opposite side of the actin filament, in which case it cannot switch to the
daughter filament, or on the same side, in which case the probability to switch
to the daughter filament is about 1/2.  Together, this gives a branching ratio
of $1/4$, not far from the exact result of $0.27$.

An experiment measuring (among other quantities) the branching ratio at Arp2/3
mediated actin filament junctions was recently carried out by
\citet{Ali.Warshaw2007}. The results are not directly comparable -- in the
experiment the actin filaments were attached to a glass surface so that some
binding sites were not accessible for myosin V.  However, because the blocked
sites are different depending whether the side filament branches to the left or
to the right, we expect that in statistical average, the calculated branching
ratio still gives a good approximation. In the experiment 18\% of the molecules
dissociated, 20\% continued on the daughter filament, and 62\% on the mother
filament. If we discount dissociation events, this means that the fraction that
switched to the daughter filament was 24\%. The statistical error of this
figure is about $\pm 5\%$ (the total number of events observed was 76).
Therefore, the result can be considered in excellent agreement with the model
calculation.

Note that our calculation only concerns stepping patterns and does not take
into account kinetics.  This has an important advantage that the result only
depends on geometric and elastic parameters, but not on kinetic constants, some
of which are still less well known \cite{Vilfan2005}.  Combining the present
model with the full kinetic scheme could influence the predicted branching
ratio as follows.  First, the binding rate of the lead head can be different in
the vicinity of the Arp2/3 complex.  And second, the ADP release rate of the
rear head can be influenced by intramolecular strain, possibly also by its
lateral (off-axis) component, as observed by Purcell and coworkers
\cite{Purcell.Spudich2005}.  Both these effects do not have a direct influence
on the branching ratio, but they might have an indirect one by influencing the
dissociation probability.  In the present calculation, events where the whole
myosin V molecule dissociates from the actin filament are not taken into
account. There is, however, evidence that the predominant dissociation path
leads through detachment of a head in the ADP state
\cite{Baker.Warshaw2004,Wu.Karplus2007} -- these processes are denoted as
Pathway 2 in Ref.~\cite{Vilfan2005}. It is therefore plausible that the
termination rate increases if either the lead head is hindered in its search
for a binding site, or the ADP release in the trail head is slowed down.  In
our model (see elastic energies in Fig.~\ref{fig:gallery}), the lead head
binding rate is strongly reduced if the trail head is bound to site $-13$, it
is also reduced somewhat if the trail head is bound to $-11$, while it is
accelerated for trail head positions $-9$ and $4$. In total, dissociation can
be accelerated by the presence of the Arp2/3 junction in 2 out of $\bar l$
cases.  Without going into quantitative details, one can conclude that the
dissociation probability could theoretically increase by up to $2/\bar l\approx
0.15$.  The branching ratio for the daughter filament could then be somewhat
smaller, because those trail head positions that have the higher branching
ratio are also more likely to lead to dissociation.  The effect caused by the
strain-dependent ADP release rate is more difficult to estimate, mainly because
the exact influence of off-axis strain on the ADP release is not yet known
quantitatively.

\begin{figure}
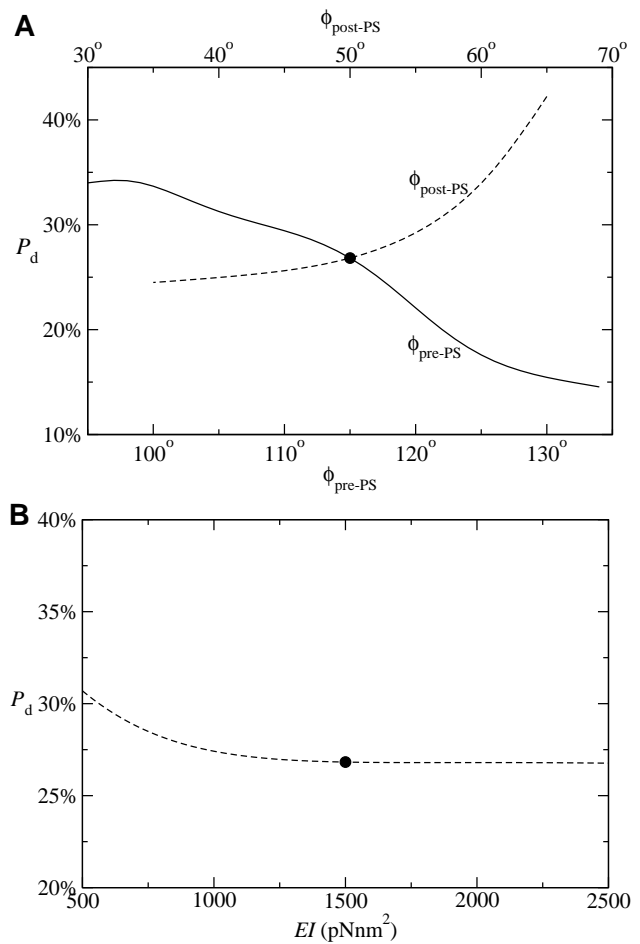

  \begin{center}
    \includegraphics{figure7a}\\
    \includegraphics{figure7b}
  \end{center}
 \caption{Dependence of the calculated branching ratio $P_d$ on different model
  parameters. (A) Dependence on the lead head angle $\phi_{\rm pre-PS}$ (solid
  line, lower scale) and trail head angle $\phi_{\rm post-PS}$ (dashed line,
  upper scale).  (B) Dependence on the lever arm stiffness (bending modulus)
  $EI$.  In both diagrams, the dot shows the value used in all other
  calculations throughout the article.}  
\label{fig:varangle}
\end{figure}

To check the robustness of our result against uncertainties in model
parameters, we calculated the variation of the branching ratio with several
model parameters.  Note that these calculations were carried out numerically
and took into account all relevant processes, including longer and shorter
steps, as well as transitions not shown in Fig.~\ref{fig:gallery}.  Data in
Fig.~\ref{fig:varangle} shows a variation of about $\pm 10\%$ if either the
lead head or the trail head angle is modified by $\pm 15^\circ$.  The allowed
variation of these two parameters that keeps the model consistent with the
experimental result is therefore restricted, although parameter sets where both
angles are increased or decreased simultaneously cannot be excluded. The result
is more robust against variations in the lever arm stiffness $EI$, where
deviations do not exceed few percent if the value of $EI$ is changed by a
factor of 3 in either direction.

We can therefore conclude that the calculated branching ratio adds support to
the elastic lever arm model presented in Ref.~\cite{Vilfan2005} and the
geometric parameters used there.  However, we cannot use it as a criterion to
determine the lever arm stiffness, which is still not known precisely.  Another
open question is to what extent the result can be reproduced with alternative
models, such as Ref.~\cite{Lan.Sun2005}, which uses a more complex model for
the elasticity of the lever arm, with a soft longitudinal (about $1/3$ of the
value used there), but very stiff azimuthal component.  The completely
different class of ``hot spot'' models, which proposes that the position of the
next binding site is determined by a propagating conformational change in the
actin filament \cite{Watanabe.Ikebe2004}, on the other hand, seems less
compatible with the finding, unless the conformational change could propagate
through the Arp2/3 complex as well.

\section*{Acknowledgements}
This work was supported by the Slovenian Research Agency (Grant P1-0099). I
would like to thank Mojca Vilfan for comments on the manuscript and Stan
Burgess for helpful discussions.

\end{document}